\documentclass[trackchanges]{aastex701}

\begin{document}
\dataset{10.5281/zenodo.20455752}   
\title{Chinese Sunspot Drawings and Their Digitization – (VIII) Release of a Machine Readable Parameters Catalog}

\author[]{Lin, G.H.}
\affiliation{State Key Laboratory of Solar Activity and Space Weather, National Astronomical Observatories, Chinese Academy of Sciences, Beijing 100101, China}
\email{lius@nao.cas.cn}
\author[0000-0002-1396-7603]{Liu, S.}
\affiliation{State Key Laboratory of Solar Activity and Space Weather, National Astronomical Observatories, Chinese Academy of Sciences, Beijing 100101, China}
\email{lius@nao.cas.cn}

\begin{abstract}
We present a systematically revised and manually verified digital catalog of historical sunspot drawing observations from Chinese observatories, encompassing sunspot number, sunspot group number, and sunspot area measurements. This represents an updated release of China's early sunspot drawings that have been utilized in numerous published studies over the years, with ongoing refinements and corrections progressively incorporated during the course of those studies and now consolidated through a systematic round of data revision aimed at improving internal consistency and usability for the broader research community. To facilitate scientific utilization, selected comparisons with international reference sunspot data are presented, focusing on relative sunspot numbers (RSN) and sunspot group numbers (GN) from Yunnan Observatory (YNO) and Purple Mountain Observatory (PMO), while sunspot area measurements are also included in this release but are not compared with international datasets in the present work. For RSN, both raw and k-corrected values are analyzed; for GN, only raw counts are used as no dedicated correction factor exists in the Chinese records. The results demonstrate strong positive correlations, with correlation coefficients consistently exceeding 0.95 for all comparisons, confirming the reliability and scientific value of these historical records. Systematic differences are observed: raw YNO RSN values are systematically higher than SIDC by about 17.6\%, while raw PMO RSN values are about 6.0\% lower. After k-correction, YNO shifts to a 20.9\% deficit, suggesting over-correction, while PMO remains lower with a 15.5\% deficit. Notably, GN exhibits smaller systematic offsets (16.5\% for YNO and 8.5\% for PMO) and serves as a more robust index for long-term solar cycle studies. This dataset serves as a valuable complement to international sunspot databases and offers unique long-term continuity for solar activity studies.
\end{abstract}

\keywords{Sunspot (1653); Solar activity(1475); Sunspot cycle(1650); Solar physics (1476)}

\section{Introduction} 
\label{sec:intro}
Historical hand-drawn sunspot drawings provide direct records of solar activity and evolution, constituting the fundamental data source for understanding long-term solar variability and its influence on the Earth environment \citep{2007AdSpR..40..929V, 2004SoPh..224...37U, 2014SSRv..186...35C}. As \citet{2014SSRv..186...35C} emphasized in their comprehensive review, the sunspot number time series is the only direct observational record at our disposal to retrace the long-term evolution of the solar cycle over the past four centuries, serving as key information in solar physics, climate studies, and space weather research. The continuous series of Chinese historical sunspot drawings constitute a unique observational dataset from the eastern hemisphere, representing an indispensable component in the global effort to reconstruct and understand solar activity patterns \citep{2019SoPh..294...79L}.

Motivated by this, we have undertaken the rescue of this regional dataset through the digitization of Chinese historical sunspot drawings. This digitization process encompassed multiple stages: selection of digitization equipment, positioning and orientation of hand-drawn recording media during scanning, determination of digitized data storage formats, intelligent recognition of physical parameters from digitized results, verification of intelligent recognition outcomes \citep{2016NewA...45...54Z, 2019SoPh..294...79L}, and ultimately the establishment of machine-readable parameter tables for six historical sunspot drawing series from six institutions or observatories across China. The scientific value of these observations has been demonstrated through numerous published studies over the years \citep{Cen1955Sunspot, 2016NewA...45...54Z, 2018IAUS..340...71Y, 2019SoPh..294...79L, 2020RAA....20...61P, 2020RAA....20..190W, 2021JApA...42...75L,2022JApA...43...41L,2021Atmos..12.1176C, 2022RAA....22i5012H}, which have progressively refined and corrected portions of the data, contributing to improved understanding of the characteristics and biases inherent in visual observations, thus enriching both Chinese and global historical sunspot activity and evolution databases.

Following the pioneering work of Wolf \citep{1861MNRAS..21...77W, 1961says.book.....W} in establishing the international sunspot number system, and the subsequent development of the group sunspot number by Hoyt and Schatten \citep{1998SoPh..179..189H, 1998SoPh..181..491H}, the importance of consistent long-term records has been widely recognized. The current release represents a systematic consolidation of these cumulative revisions, incorporating a comprehensive round of data verification and correction aimed at improving internal consistency and usability for the broader research community. Following best practices in historical data management, we have applied rigorous quality control procedures and documented all modifications to ensure transparency and reproducibility. Future updates and refinements are anticipated as part of ongoing efforts to enhance data quality.

The Chinese sunspot drawing dataset encompasses multiple physical parameters, including sunspot number, sunspot group number, sunspot area measurements, sunspot distribution, and extensive observation information, as detailed in \citep{2019SoPh..294...79L, 2020RAA....20...61P}. To facilitate scientific utilization of this comprehensive dataset, we present selected comparisons with international reference data from the Solar Influences Data Analysis Center (SIDC: \url{https://www.sidc.be/SILSO/home}) \citep{1987A&AS...70...83W, 2004SoPh..224..113V, 2014SSRv..186...35C, 2016SoPh..291.2629C,  2019NatAs...3..205M, 2023SoPh..298...44C}, focusing specifically on sunspot number and sunspot group number from Yunnan Observatory (YNO) and Purple Mountain Observatory (PMO). Sunspot area measurements and distribution information are also included in this release, although they are not compared with international datasets in the present work. These comparisons are intended to assess the consistency between Chinese observations and international standards, and to characterize any systematic differences that may arise from instrumental, observational, or geographical factors \citep{2001MNRAS.323..223B, 2016SoPh..291.3081B}. Such cross-comparisons are essential for understanding the place of regional datasets within the global framework of solar activity monitoring.

This paper is organized as follows: Section 2 describes the Chinese sunspot drawing data sources. Section 3 presents comparative analyses between YNO and PMO observations and international reference data. Section 4 states the main conclusions, followed by brief discussions.
\section{Sunspot Drawing Data Sources}
\label{sec:observation}

China's historical sunspot drawings originate from six astronomical institutions distributed across the country. These stations, with their varying geographical locations and observation periods, collectively provide nearly a century of solar records from the eastern hemisphere.
Among these six stations, Yunnan Observatory and Purple Mountain Observatory stand out for their exceptional data continuity and volume (YNO: 1957-; PMO: 1954-2011), together contributing the majority of records to our archive. The remaining four stations, while having more fragmented observation histories, offer valuable supplementary data that fills specific temporal gaps. The digitized data for all stations are publicly available at \url{https://sun10.bao.ac.cn/hsos_data/SHDA/}.

\subsection{Yunnan Astronomical Observatory}

Located in Kunming at geographic coordinates \(25^\circ01'\) N, \(102^\circ47'\) E, Yunnan Observatory holds the most comprehensive and continuous sunspot drawing record in China. Its observation, initiated in 1957 and continuing to the present day, has accumulated over six decades of daily sunspot drawings. The instrumental setup consists of an equatorial refractor with 12.7 cm aperture and 195 cm focal length, configured to project a solar image \textbf{with a diameter of 17.4 cm} radius onto recording paper. This standardized projection diameter, adopted nationally in 1957, ensures consistency across different stations and time periods. The YNO collection encompasses approximately 15,800 individual drawings, from which our digitization efforts have extracted nearly 90,000 handwritten records and over one million discrete information entries. The scientific merit of this dataset has been acknowledged in comparative studies \citep{2001MNRAS.323..223B}, which noted the exceptionally low random errors in YNO's sunspot area measurements, making them particularly valuable for complementing observations from other longitudes.

\subsection{Purple Mountain Astronomical Observatory}

Purple Mountain Observatory (\(32^\circ03'\) N, \(118^\circ49'\) E) in Nanjing holds historical significance as the institution where systematic sunspot recording methodology was standardized across China. Following wartime disruptions, PMO resumed regular observations in 1954 and subsequently established the national protocol requiring a uniform 17.4 cm solar image diameter. The primary instrument employed was a 20 cm Zeiss equatorial refractor with 350 cm focal length. Over its 57-year operational span (1954–2011, with some intermittent gaps), PMO amassed approximately 12,500 sunspot drawings. These records contain over 57,000 handwritten entries, translating to more than 626,000 individual data points in our digital archive. The PMO series holds particular value not only for its temporal extent but also because its observational practices served as the observation template specification for other Chinese stations, ensuring methodological consistency across the national network. 

\subsection{Complementary Stations}

Four additional observatories contribute to the completeness of China's sunspot drawing heritage.

\textbf{Qingdao Observatory Guang Xiang Station (QDGX)} at \(36^\circ04'\) N, \(120^\circ19'\) E holds the distinction of producing China's first telescopic sunspot drawing on 1 May 1925, initiated by astronomer Gao Pingzi using a 16 cm equatorial telescope. Despite multiple interruptions—including the 1937 invasion and subsequent social upheavals—the station accumulated over 12,600 drawings beginning from 1925.

\textbf{Sheshan Observing Station (SSOS)} at \(31^\circ06'\) N, \(121^\circ13'\) E contributed more than 2,400 drawings during 1952–1964.

\textbf{Beijing Planetarium (BJP)} at \(39^\circ54'\) N, \(116^\circ23'\) E provided approximately 4,000 drawings spanning two periods (1979–1982 and 1989–1999).

\textbf{Nanjing University (NJU)} at \(32^\circ03'\) N, \(118^\circ51'\) E added observations from 1986–2002 and 2004–2015, contributing approximately 3,500 drawings.

Though these stations individually offer shorter or more fragmented records, their collective contribution ensures that China's sunspot observations achieve near-continuous coverage from 1947 onward, effectively filling the longitudinal gap in global solar monitoring networks.

The digitized parameters from all six observatories are publicly available in machine-readable format; see the Data Availability section for access information.

\subsection{Example of Sunspot Drawing and Digitized Parameters}

\begin{figure}[htbp]
    \centering
    \includegraphics[width=\linewidth]{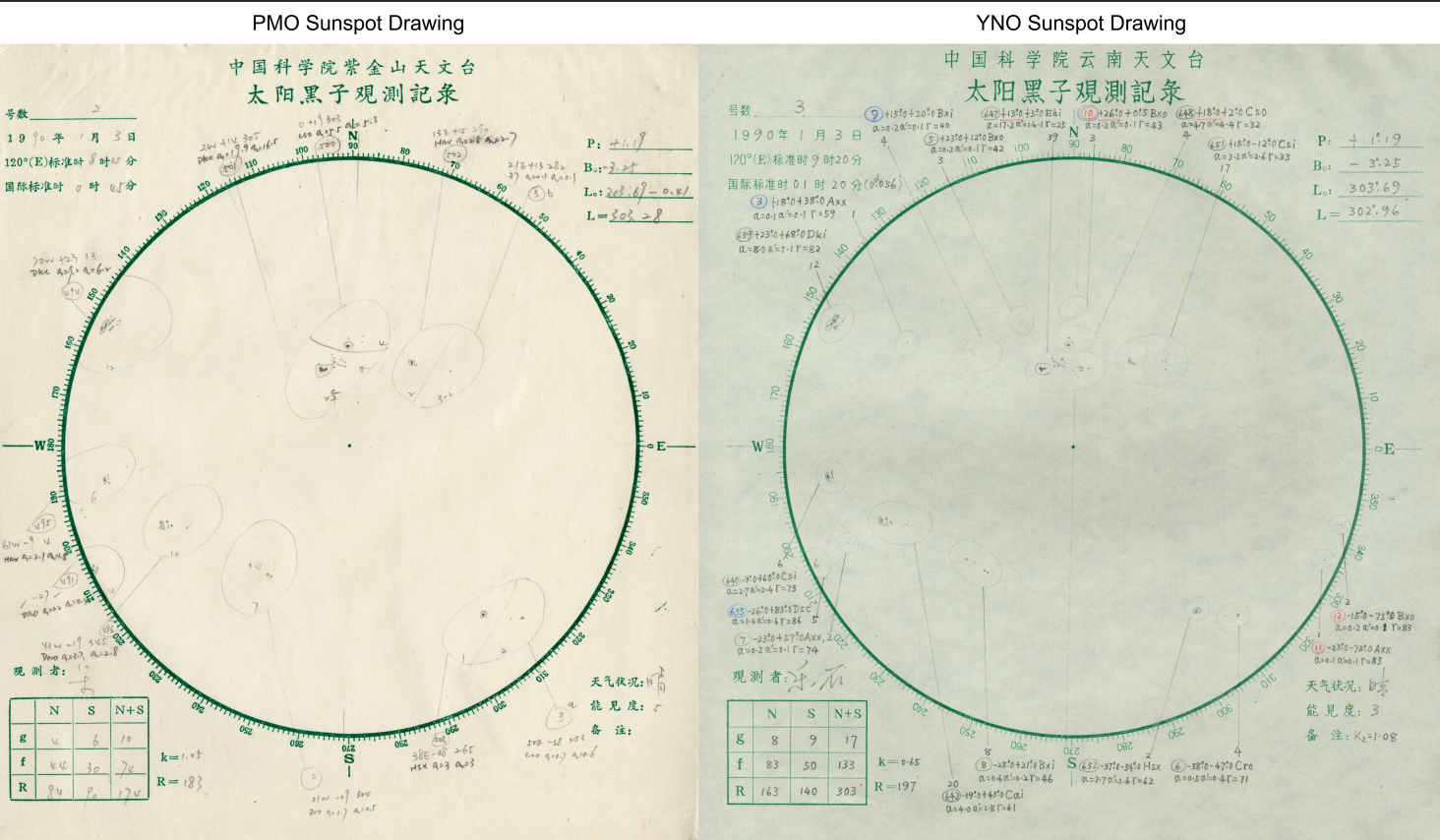}
    \caption{The sunspot drawings from PMO (left) and YNO (right) on January 3, 1990. The PMO drawing was observed at 00:45 UTC, and the YNO drawing at 01:20 UTC. Both drawings follow the same standardized recording template, yet differences in sunspot depiction reflect variations in local seeing conditions, instrumentation, and individual observer practices.}
    \label{fig:PMO_YNO_comparison}
\end{figure}

Figure~\ref{fig:PMO_YNO_comparison} shows representative sunspot drawings from PMO and YNO observed on January 3, 1990, illustrating the original sunspot drawing observations from two different stations on the same day. Despite being observed on the same date and following consistent recording templates, noticeable differences exist between the two drawings. These variations can be attributed to various factors including local observing conditions, telescope specifications, observer experience, and the subjective nature of manual sunspot depiction. The consistent template design across stations ensures that the recorded information follows a uniform structure, facilitating systematic digitization and comparison, while the inherent differences highlight the importance of cross-calibration and careful quality assessment when combining data from multiple stations.

After scanning, we performed information extraction from the digitized images. Following the description in \citet{2019SoPh..294...79L}, the raw data recorded in each sunspot drawing can be categorized into two types: fixed-format information and non-fixed information.

The \textbf{fixed-format information} is recorded in pre-printed rectangular boxes and lines at specific positions on the observation paper, containing standardized metadata. While the overall structure is consistent across stations, minor variations exist in the specific parameters recorded due to different observational practices and historical developments.

For PMO observations, the fixed-format information includes: sequential number (NUM), observation date (DATE), Beijing time (CST), Universal Time Coordinated (UTC), observation day count (DAYS), heliographic coordinates (P, B0, L0, L1, L), sunspot group counts in the northern hemisphere (gN), southern hemisphere (gS), and total (gNS), sunspot counts in the northern hemisphere (fN), southern hemisphere (fS), and total (fNS), Wolf numbers for the northern hemisphere ((10g+f)N), southern hemisphere ((10g+f)S), and total ((10g+f)NS), Wolf numbers corrected by the k factor (SpN, SpS, SpNS), corrected coefficients (k for sunspots, k2 for sunspot area), the final relative sunspot number corrected by k (R), visible sunspot count (Visible), as shown in Table~\ref{tab:fixed_format}, \textbf{where missing entries are left blank.}

For YNO observations, the fixed-format information includes similar parameters but with some differences in notation and structure: sequential number (NUM), observation date (DATE), Beijing time (CST), Universal Time Coordinated (UTC), observation day count (days), heliographic coordinates (P, B0, L0, L), sunspot group counts in the northern hemisphere (gN), southern hemisphere (gS), and total (gNS), sunspot counts in the northern hemisphere (fN), southern hemisphere (fS), and total (fNS), Wolf numbers for the northern hemisphere (RN), southern hemisphere (RS), and total (RNS), corrected coefficients (k for sunspots, k2 for sunspot area), the final relative sunspot number corrected by k (R), and visible sunspot count (Visible), as shown in Table~\ref{tab:yno_fixed_format}. Notably, YNO records use RN, RS, RNS to denote Wolf numbers instead of the (10g+f)N/S/NS notation used by PMO, and do not include the L1 coordinate or the SpN/SpS/SpNS area parameters present in PMO records.

For sunspot number and sunspot group related parameters, the tables from different stations exhibit good overall consistency. However, minor variations exist across different stations and different observing periods. In this work, we have carefully reviewed and upgraded these tables, providing a unified dataset that harmonizes records from all stations and periods while removing redundant information, such as the CST, gNS=gN+gS, fNS=fN+fS. The standardized data format ensures consistency and facilitates comparative analysis across the entire Chinese sunspot drawing archive.

Table~\ref{tab:yno_unified} shows an example of the unified format for sunspot number and sunspot group parameters after standardization. This unified table retains the essential physical parameters while eliminating the redundant fields, enabling direct comparison and combined analysis of observations from different stations and time periods.

\begin{table}[htbp]
    \centering
\tiny
    \setlength{\tabcolsep}{2pt}
    \caption{Example of fixed-format information extracted from PMO sunspot drawings (January 2-8, 1990).}
    \label{tab:fixed_format}
    \begin{tabular}{*{26}{c}}
        \hline
        NUM & DATE & CST & UTC & DAYS & P & B0 & L0 & L1 & L & gN & gS & gNS & fN & fS & fNS & (10g+f)N & (10g+f)S & (10g+f)NS & SpN & SpS & SpNS & k & R & Visible & k2 \\
        \hline
        1 & 19900102 & 835 & 35 & & 1.67 & -3.14 & 316.86 & 0.32 & 316.54 & 4 & 5 & 9 & 29 & 27 & 56 & 69 & 77 & 146 & & & & 1.05 & 153 & 3 & \\
        2 & 19900103 & 845 & 45 & & 1.19 & -3.25 & 303.69 & 0.41 & 303.28 & 4 & 6 & 10 & 44 & 30 & 74 & 84 & 90 & 174 & & & & 1.05 & 183 & 5 & \\
        3 & 19900104 & 845 & 45 & & 0.70 & -3.37 & 290.52 & 0.41 & 290.11 & 4 & 6 & 10 & 21 & 19 & 40 & 61 & 79 & 140 & & & & 1.05 & 147 & 5 & \\
        4 & 19900105 & 810 & 10 & & 0.22 & -3.48 & 277.35 & 0.09 & 277.26 & 4 & 4 & 8 & 33 & 10 & 43 & 73 & 50 & 123 & & & & 1.05 & 129 & 4 & \\
        5 & 19900106 & 830 & 30 & & -0.27 & -3.59 & 264.18 & 0.28 & 263.90 & 5 & 5 & 10 & 25 & 15 & 40 & 75 & 65 & 140 & & & & 1.05 & 147 & 5 & \\
        6 & 19900107 & 855 & 55 & & -0.75 & -3.70 & 251.01 & 0.51 & 250.50 & 5 & 4 & 9 & 29 & 14 & 43 & 79 & 54 & 133 & & & & 1.05 & 140 & 5 & \\
        7 & 19900108 & 1400 & 600 & & & & & & & & & & & & & & & & & & & 1.05 & & 1 & \\
        \hline
    \end{tabular}
\end{table}
\begin{table}[htbp]
    \centering
\tiny
    \setlength{\tabcolsep}{2pt}  
    \caption{Example of fixed-format information extracted from YNO sunspot drawings (January 1-10, 1990).}
    \label{tab:yno_fixed_format}
    \begin{tabular}{*{25}{c}}
        \hline
        NUM & DATE & CST & UTC & days & P & B0 & L0 & L & gN & gS & gNS & fN & fS & fNS & RN & RS & RNS & k & R & Visible & k2 \\
        \hline
        1 & 19900101 & 950 & 150 & 0.077 & 2.16 & -3.02 & 330.03 & 329.02 & 8 & 8 & 16 & 72 & 69 & 141 & 152 & 149 & 301 & 0.65 & 196 & 3 & 1.08 \\
	   2 & 19900102 & 1020 & 220 & 0.097 & 1.67 & -3.14 & 316.86 & 315.58 & 8 & 8 & 16 & 70 & 58 & 128 & 150 & 138 & 288 & 0.65 & 187 & 2 & 1.08 \\
        3 & 19900103 & 920 & 120 & 0.056 & 1.19 & -3.25 & 303.69 & 302.96 & 8 & 9 & 17 & 83 & 50 & 133 & 163 & 140 & 303 & 0.65 & 197 & 3 & 1.08 \\
        4 & 19900104 & 950 & 150 & 0.077 & 0.70 & -3.37 & 290.52 & 289.51 & 7 & 9 & 16 & 64 & 51 & 115 & 134 & 141 & 275 & 0.65 & 179 & 4 & 1.08 \\
        5 & 19900105 & 905 & 105 & 0.045 & 0.22 & -3.48 & 277.35 & 276.75 & 6 & 9 & 15 & 88 & 34 & 122 & 148 & 124 & 272 & 0.65 & 177 & 3 & 1.08 \\
	   6 & 19900106 & 930 & 130 & 0.063 & -0.27 & -3.59 & 264.18 & 263.35 & 6 & 7 & 13 & 65 & 27 & 92 & 125 & 97 & 222 & 0.65 & 144 & 3 & 1.08 \\
        7 & 19900107 & 940 & 140 & 0.070 & -0.75 & -3.70 & 251.01 & 250.09 & 5 & 6 & 11 & 45 & 38 & 83 & 95 & 98 & 193 & 0.64 & 124 & 3 & 1.01 \\
        8 & 19900108 & 940 & 140 & 0.070 & -1.23 & -3.81 & 237.84 & 236.92 & 5 & 6 & 11 & 19 & 46 & 65 & 69 & 106 & 175 & 0.64 & 112 & 2 & 1.01 \\
        9 & 19900109 & 1010 & 210 & 0.090 & -1.71 & -3.92 & 224.67 & 223.48 & 4 & 6 & 10 & 13 & 51 & 64 & 53 & 111 & 164 & 0.64 & 105 & 2 & 1.01 \\
        10 & 19900110 & 930 & 130 & 0.063 & -2.19 & -4.03 & 211.50 & 210.67 & 3 & 8 & 11 & 18 & 63 & 81 & 48 & 143 & 191 & 0.64 & 122 & 3 & 1.01 \\
        \hline
    \end{tabular}
\end{table}

\begin{table}[htbp]
    \centering
    \caption{Example of unified format for sunspot parameters from YNO observations (January 1-10, 1990).}
    \label{tab:yno_unified}
    \begin{tabular}{cccccccccccc}
        \hline
        DATE & UTC & gN & gS & gNS(g) & fN & fS & fNS(f) & k & Visible & k2 \\
        \hline
        19900101 & 150 & 8 & 8 & 16 & 72 & 69 & 141 & 0.65 & 3 & 1.08 \\
        19900102 & 220 & 8 & 8 & 16 & 70 & 58 & 128 & 0.65 & 2 & 1.08 \\
        19900103 & 120 & 8 & 9 & 17 & 83 & 50 & 133 & 0.65 & 3 & 1.08 \\
        19900104 & 150 & 7 & 9 & 16 & 64 & 51 & 115 & 0.65 & 4 & 1.08 \\
        19900105 & 105 & 6 & 9 & 15 & 88 & 34 & 122 & 0.65 & 3 & 1.08 \\
        19900106 & 130 & 6 & 7 & 13 & 65 & 27 & 92 & 0.65 & 3 & 1.08 \\
        19900107 & 140 & 5 & 6 & 11 & 45 & 38 & 83 & 0.64 & 3 & 1.01 \\
        19900108 & 140 & 5 & 6 & 11 & 19 & 46 & 65 & 0.64 & 2 & 1.01 \\
        19900109 & 210 & 4 & 6 & 10 & 13 & 51 & 64 & 0.64 & 2 & 1.01 \\
        19900110 & 130 & 3 & 8 & 11 & 18 & 63 & 81 & 0.64 & 3 & 1.01 \\
        \hline
    \end{tabular}
\end{table}

The \textbf{non-fixed information} is recorded in handwritten characters about individual sunspots and sunspot groups, with their positions varying according to the specific sunspot groups being described. For YNO observations during the period shown in Table~\ref{tab:yno_nonfixed_format}, the information is remarkably rich, a level of detail rarely found in early historical records. It typically includes: date, sunspot group number, heliographic longitude and latitude, McIntosh classification type, umbra area, penumbra area, the linear distance (r) between the center of mass of the sunspot group and the center of the solar disk, and the number of sunspots within the group.

It should be noted that this represents only one example from a specific observing period. The format, content, and notation of non-fixed information vary considerably across different stations, and even within the same station over different observing periods. Some parameters may be present in certain records while absent in others, and the symbols used to represent the same physical parameters may differ. Therefore, the specific information recorded in the non-fixed format should be interpreted in conjunction with the detailed documentation accompanying each data table for the corresponding observing period and station.
\begin{table}[htbp]
    \centering
    \small
    \setlength{\tabcolsep}{3pt}
    \caption{Example of non-fixed information extracted from YNO sunspot drawings (January 1-3, 1990).}
    \label{tab:yno_nonfixed_format}
    \begin{tabular}{cccccccccc} \hline
        DATE & NUM & LAT & LON & TYPE & Umbra area & Penumbra area & radius & Number of sunspots \\
        \hline
        19900101 & 1 & -16 & 24 & BXO & 0.2 & 0.1 & 40 & 2 \\
        19900101 & 2 & 14 & 15 & AXX & 0.2 & 0.1 & 34 & 2 \\
        19900101 & 3 & 23 & 13 & AXX & 0.2 & 0.1 & 43 & 4 \\
        19900101 & 4 & 13 & -13 & AXX & 0.2 & 0.1 & 30 & 2 \\
        19900101 & 5 & 24 & -20 & AXX & 0.2 & 0.1 & 47 & 2 \\
        19900101 & 6 & -38 & -70 & BXO & 0.2 & 0.1 & 83 & 3 \\
        19900101 & 635 & -26 & 59 & EKC & 24.5 & 20.0 & 75 & 27 \\
        19900101 & 636 & -11 & 51 & AXX & 0.2 & 0.2 & 67 & 1 \\
        19900101 & 639 & 23 & 42 & DKC & 20.6 & 15.3 & 65 & 16 \\
        19900101 & 640 & -9 & 34.5 & CSI & 6.7 & 5.8 & 49 & 14 \\
        19900101 & 642 & -18 & 18 & DSI & 5.9 & 5.1 & 34 & 20 \\
        19900101 & 647 & 13 & -22 & EKI & 16.8 & 13.9 & 39 & 22 \\
        19900101 & 648 & 18 & -24 & DSI & 4.8 & 4.1 & 49 & 4 \\
        19900101 & 650 & -17 & 2 & AXX & 0.1 & 0.1 & 21 & 1 \\
        19900101 & 651 & 15 & -40 & CAI & 2.6 & 2.2 & 59 & 20 \\
        19900101 & 652 & -37 & -58 & HSX & 1.5 & 1.5 & 77 & 1 \\
        19900102 & 2 & 14 & 29.5 & AXX & 0.1 & 0.1 & 48 & 1 \\
        19900102 & 3 & 19 & 25 & AXX & 0.1 & 0.1 & 47 & 1 \\
        19900102 & 5 & 24 & -7 & BXI & 0.2 & 0.1 & 40 & 5 \\
        19900102 & 6 & -38 & -58 & AXX & 0.2 & 0.1 & 77 & 2 \\
        19900102 & 7 & -25 & 45 & BXO & 0.2 & 0.1 & 65 & 5 \\
        19900102 & 8 & -28 & 9.5 & BXO & 0.2 & 0.1 & 39 & 5 \\
        19900102 & 635 & -26 & 72 & EKC & 11.3 & 8.1 & 83 & 14 \\
        19900102 & 639 & 23 & 56 & DKI & 11.3 & 10.2 & 75 & 13 \\
        19900102 & 640 & -9 & 47.5 & CSI & 4.6 & 4.0 & 64 & 7 \\
        19900102 & 642 & -18 & 31 & DSI & 4.5 & 3.6 & 48 & 22 \\
        19900102 & 647 & 13 & -8 & EHI & 17.5 & 14.8 & 27 & 24 \\
        19900102 & 648 & 19 & -11 & CSI & 5.6 & 4.7 & 35 & 7 \\
        19900102 & 650 & -17 & 17 & AXX & 0.1 & 0.1 & 33 & 1 \\
        19900102 & 651 & 15 & -25 & CSI & 3.2 & 2.5 & 44 & 18 \\
        19900102 & 652 & -37 & -45 & HSX & 1.9 & 1.8 & 70 & 2 \\
        19900102 &   & 17 & 9 & AXX & 0.1 & 0.1 & 32 & 1 \\
        19900103 & 3 & 18 & 38 & AXX & 0.1 & 0.1 & 59 & 1 \\
        19900103 & 5 & 23 & 12 & BXO & 0.2 & 0.1 & 42 & 3 \\
        19900103 & 6 & -38 & -47 & CRO & 0.5 & 0.4 & 71 & 4 \\
        19900103 & 9 & 15 & 20 & BXI & 0.2 & 0.1 & 40 & 4 \\
        19900103 & 10 & 26 & 0.5 & BXO & 0.2 & 0.1 & 43 & 3 \\
        19900103 & 11 & -23 & -73 & AXX & 0.1 & 0.1 & 83 & 1 \\
        19900103 & 12 & -15 & -75 & BXO & 0.2 & 0.1 & 83 & 2 \\
        19900103 & 647 & 13 & 5 & EHI & 17.2 & 14.1 & 25 & 39 \\
        19900103 & 648 & 18 & 2 & CSO & 4.7 & 4.4 & 32 & 4 \\
        19900103 & 651 & 15 & -12 & CSI & 3.2 & 2.6 & 33 & 17 \\
        19900103 & 652 & -37 & -34 & HSX & 2.7 & 2.6 & 62 & 2 \\
        \hline
    \end{tabular}
\end{table}

\section{Comparison between YNO/PMO and SIDC Data}
\label{sec:comparison}
\subsection{Comparison Method}
To evaluate the quality and consistency of the digitized Chinese sunspot drawings, it is essential to compare them with internationally recognized reference data. Such comparisons not only validate the reliability of our dataset but also help characterize systematic differences that may arise from instrumental, observational, or geographical factors. The Solar Influences Data Analysis Center (SIDC) provides the most comprehensive and authoritative sunspot reference data, including the international sunspot number and sunspot group number series.
The SIDC reference data used in this study are publicly available at \url{https://www.sidc.be/SILSO/home}. 

Following the classical definition introduced by Wolf, the relative sunspot number (RSN) is calculated as:
\begin{equation}
RSN = k (10g + f)
\end{equation}
where \(g\) is the number of sunspot groups, \(f\) is the number of individual sunspots, and \(k\) is a corrected coefficient that accounts for differences in observing conditions, instruments, and observer experience. Specifically, we utilize the daily total sunspot number series (SN\_d\_tot\_V2.0) for RSN comparisons.

\textbf{For sunspot group numbers (GN), we use the SILSO reconstructed dataset series: the daily backbone series \citep{1998SoPh..179..189H, 1998SoPh..181..491H} GNhs\_d.txt (1610–1995) for daily comparisons, and the annual backbone series \citep{2016SoPh..291.2653S} GNbb2\_y.txt (1610–2015) for annual comparisons. Since the daily reconstructed GN series ends in 1995, we complement it with the annual series to cover the full time range of each observatory.}

\textbf{For each comparison, we perform ordinary least squares (OLS) linear regression with SIDC values as the dependent variable and the observatory values as the independent variable, following the form:}

\begin{equation}
\text{SIDC} = \text{slope} \times \text{Obs} + \text{intercept}
\end{equation}

\textbf{where Obs represents either YNO or PMO measurements (sunspot number or group number). For sunspot numbers (RSN), we analyze both raw and k-corrected values to assess the impact of the correction factor. For group numbers (GN), we analyze only the raw counts, as our Chinese records do not contain a dedicated correction factor for groups (k\_G). The resulting slope quantifies the systematic offset between the datasets, with slopes less than unity indicating that SIDC values are systematically lower than the observatory measurements, and slopes greater than unity indicating the opposite. The correlation coefficient 
$r$ provides a measure of the strength of the linear relationship between the two datasets. The intercept quantifies any constant offset between the two datasets when the observatory value is zero.}

\subsection{YNO Sunspot Observation Comparison}

\subsubsection{Relative Sunspot Number (RSN) Analysis for YNO}
\begin{figure}[htbp]
    \centering
    \includegraphics[width=\linewidth]{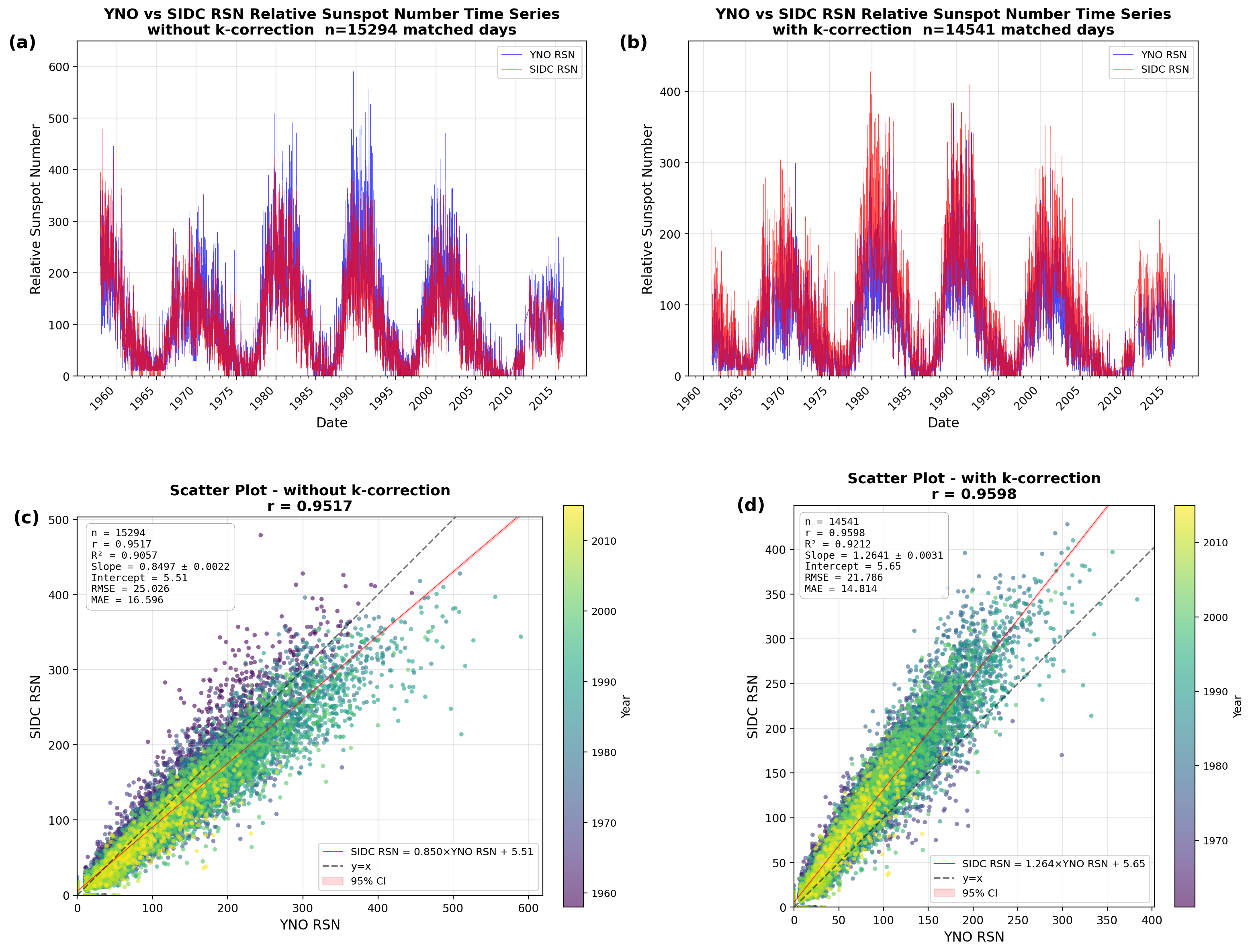}
    \caption{Comparison between YNO relative sunspot numbers (RSN) and SIDC reference data. The top row shows time series comparison; the bottom row shows scatter plots with regression analysis. Raw (left column) and k-corrected (right column) comparisons are presented. Legend boxes display statistical parameters including sample size $n$, correlation coefficient $r$, coefficient of determination $R^2$, regression slope and intercept, RMSE, and MAE.}
    \label{fig:YNO_SIDC_RSN}
\end{figure}
Figure~\ref{fig:YNO_SIDC_RSN} shows the comparison between YNO relative sunspot numbers (RSN) and SIDC reference data. For RSN, both raw (without k-correction) and k-corrected (with k-correction) values are presented, including time series (panels a, b) and scatter plots with regression analysis (panels c, d). The scatter plots include regression lines, 95\% confidence intervals, and legend boxes containing key statistical parameters ($n$, $r$, $R^2$, slope, intercept, RMSE, MAE).

\begin{table}[htbp]
\centering
\caption{Statistical summary of YNO RSN vs. SIDC comparison}
\label{tab:yno_comparison_stats_rsn}
\begin{tabular}{lcccccccc}
\hline
Dataset & $n$ & $r$ & $R^2$ & Slope & Intercept & SEE & RMSE & MAE \\
\hline
YNO RSN (raw) & 15,294 & 0.9517 & 0.9057 & $0.8497\pm0.0022$ & $5.507\pm0.316$ & 25.03 & 25.03 & 16.60 \\
YNO RSN (k) & 14,541 & 0.9598 & 0.9212 & $1.2641\pm0.0031$ & $5.649\pm0.279$ & 21.79 & 21.79 & 14.81 \\
\hline
\end{tabular}
\end{table}

The statistical results for YNO RSN are summarized in Table~\ref{tab:yno_comparison_stats_rsn}. The correlation coefficient $r$ measures the strength of the linear relationship between YNO and SIDC values. The slope indicates systematic offset: slope $<1$ means YNO values are higher than SIDC, slope $>1$ means YNO values are lower than SIDC. The SEE (standard error of the estimate), RMSE (root mean square error), and MAE (mean absolute error) are error metrics that quantify prediction accuracy — smaller values indicate better agreement between YNO and SIDC.

For the raw YNO RSN (without k-correction), the regression analysis against SIDC reference data yields a slope of $0.8497 \pm 0.0022$ and an intercept of $5.507 \pm 0.316$. The correlation coefficient is $r = 0.9517$ ($R^2 = 0.9057$), with $n = 15,294$ matched daily observations. The error metrics are SEE = 25.03, RMSE = 25.03, and MAE = 16.60. The slope of 0.850 indicates that raw YNO RSN values are systematically higher than SIDC values. The 95\% confidence interval for the slope is $[0.8453, 0.8540]$, which does not include unity, confirming that this offset is statistically significant.

For the k-corrected YNO RSN ($n = 14,541$ matched days), the k-correction factors range from 0.510 to 0.980 with a mean value of 0.656. After applying k-correction, the regression slope becomes $1.2641 \pm 0.0031$, with an intercept of $5.649 \pm 0.279$. The correlation coefficient improves to $r = 0.9598$ ($R^2 = 0.9212$). All three error metrics decrease: SEE from 25.03 to 21.79, RMSE from 25.03 to 21.79, and MAE from 16.60 to 14.81, indicating that k-correction improves the prediction accuracy. The slope of 1.264 indicates that k-corrected YNO RSN values are systematically lower than SIDC values. The 95\% confidence interval $[1.2581, 1.2701]$ does not include unity, confirming the statistical significance of this offset.

\subsubsection{Residual Analysis of RSN for YNO}

\begin{figure}[htbp]
    \centering
    \includegraphics[width=\linewidth]{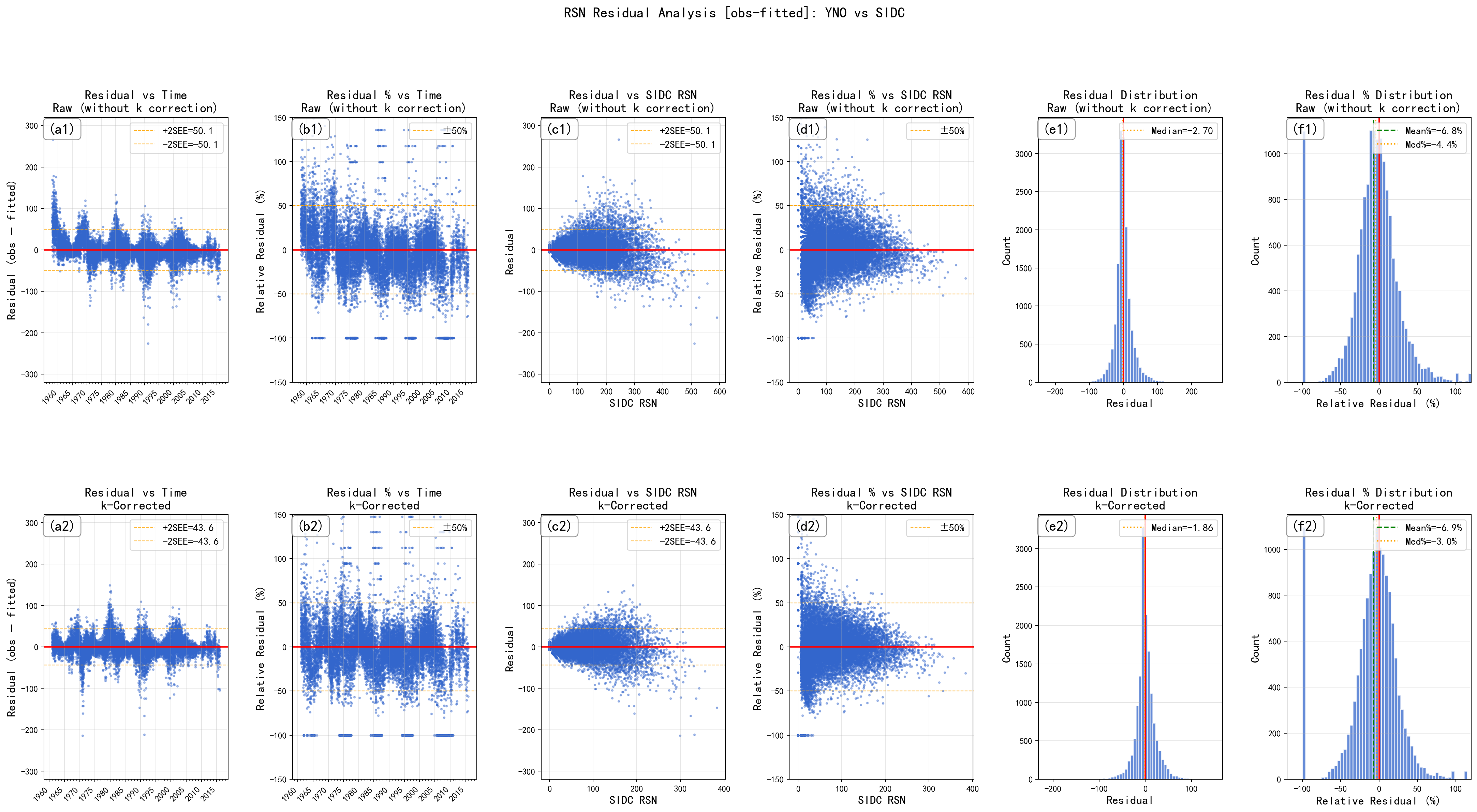}
    \caption{Residual analysis for YNO RSN (obs $-$ fitted). Top row: raw data; bottom row: k-corrected data. Columns show (1) residuals vs time, (2) percentage residuals vs time, (3) residuals vs reference SIDC RSN, (4) percentage residuals vs reference, (5) residual histogram, (6) percentage residual histogram. Dashed lines indicate $\pm2$SEE (orange) and $\pm50\%$ (orange).}
    \label{fig:rsn_yno_residual_fitted}
\end{figure}

Figure~\ref{fig:rsn_yno_residual_fitted} presents the residual analysis for YNO RSN based on the regression model (residual = observed $-$ fitted). Here, fitted values are the SIDC values predicted by the regression line ($\text{fitted} = \text{slope} \times \text{YNO}_{\text{obs}} + \text{intercept}$). The residual thus measures the deviation of YNO observations from the regression line. The figure consists of two rows (raw data and k-corrected data) and six columns, showing different diagnostic views.

The first two columns (Col 1-2) show residuals and percentage residuals versus time. For raw YNO RSN, residuals are scattered around zero with a standard error of the estimate (SEE) of 25.03. The $\pm2$SEE boundaries ($\pm50.05$) capture most data points. For k-corrected data, the SEE reduces to 21.79, indicating improved precision after correction, with residuals more tightly clustered around zero.

The absolute residuals (Col 1) exhibit a clear solar cycle modulation: larger residuals appear around solar maxima when sunspot activity is high, while residuals are smaller during solar minima. This pattern is expected, as higher activity levels naturally lead to larger absolute deviations. For percentage residuals (Col 2, residual/fitted $\times$ 100\%), the opposite trend is observed — relative errors tend to be larger during solar minima when fitted values are small, though this trend is not strongly pronounced. This behavior is typical in sunspot data analysis, as the denominator becomes very small during quiet periods, amplifying the relative error even for small absolute deviations. The scatter plots (Col 3-4), which show residuals and percentage residuals versus reference SIDC values, confirm these patterns: absolute residuals (Col 3) increase with reference RSN values, while percentage residuals (Col 4) show a slight decrease with activity level, with larger scatter at low activity levels (small RSN). After k-correction, both absolute and percentage residuals show reduced overall scatter, consistent with the improved SEE and RMSE values in Table~\ref{tab:yno_comparison_stats_rsn}.

The histograms (Col 5-6) show the distribution of residuals and percentage residuals. For raw YNO RSN (Col 5), the residual distribution is centered near zero with a median of $-2.70$. For k-corrected data, the median improves to $-1.86$, consistent with the reduced RMSE (from 25.03 to 21.79). The percentage residual histograms (Col 6) show that for raw data, the mean percentage residual is $-6.8\%$ with a median of $-4.4\%$. After k-correction, the mean percentage residual becomes $-6.9\%$ with a median of $-3.0\%$. The overall spread decreases, confirming that k-correction reduces both absolute and relative prediction errors.

\subsubsection{Group Number (GN) Analysis for YNO}

\begin{figure}[htbp]
    \centering
    \includegraphics[width=\linewidth]{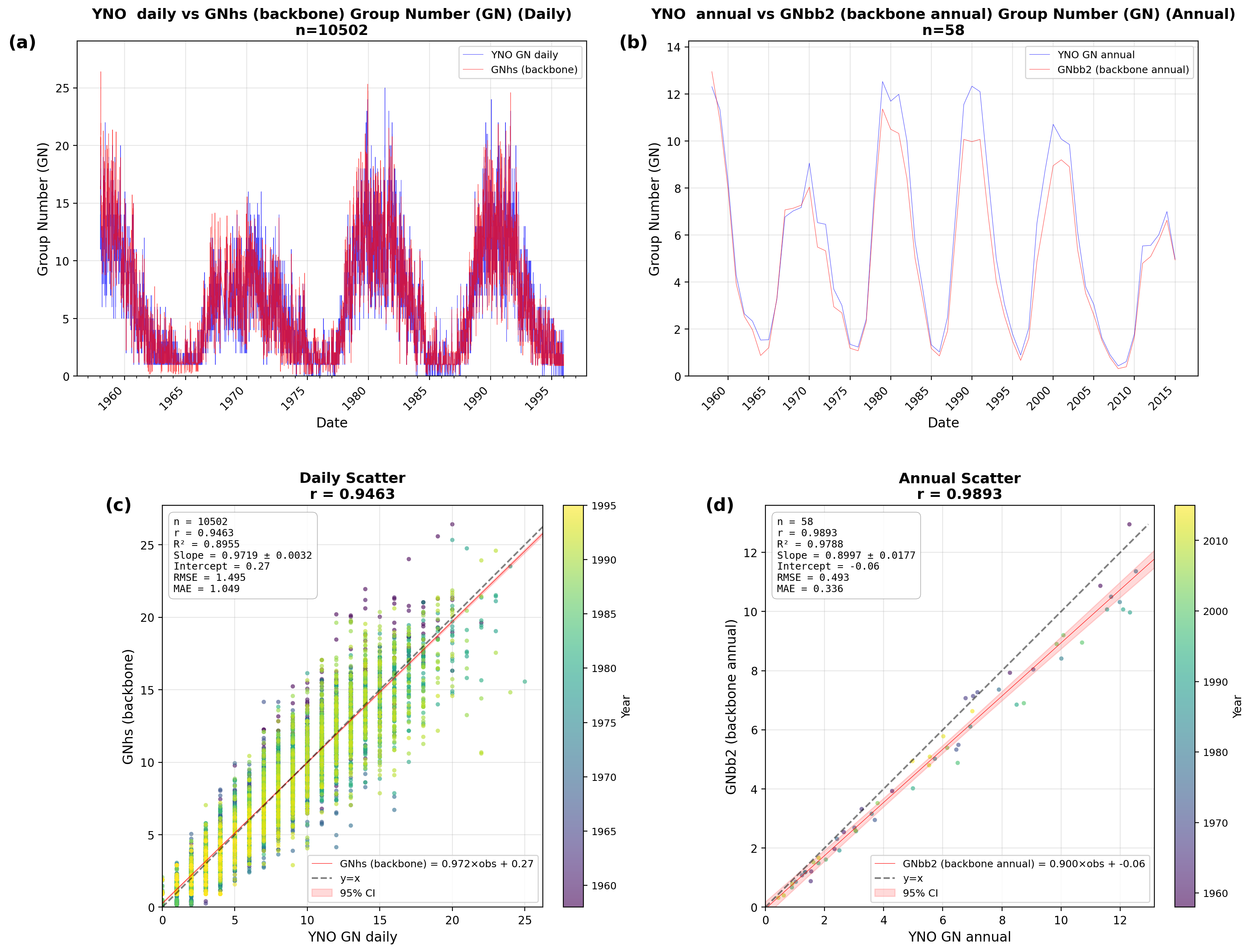}
    \caption{Comparison between YNO group numbers (GN) and SIDC reconstructed GN series (GNhs daily and GNbb2 annual). The top row shows time series comparisons; the bottom row shows scatter plots with regression analysis. Daily comparison (left column) uses GNhs\_d.txt (1958–1995); annual comparison (right column) uses GNbb2\_y.txt (1958–2015). Legend boxes display the same statistical parameters as in Figure~\ref{fig:YNO_SIDC_RSN}.}
    \label{fig:YNO_SIDC_GN}
\end{figure}

Figure~\ref{fig:YNO_SIDC_GN} shows the comparison for group numbers (GN), with daily and annual comparisons presented separately. The daily comparison (1958–1995) uses the SILSO reconstructed daily series (GNhs\_d.txt), while the annual comparison (1958–2015) uses the annual backbone series (GNbb2\_y.txt). The annual GN values for YNO are obtained by averaging daily observations within each year. The statistical results are summarized in Table~\ref{tab:yno_comparison_stats_gn}. As described in the RSN analysis, the correlation coefficient $r$ quantifies linear agreement, the slope indicates systematic offset (slope $<1$: YNO higher than SIDC; slope $>1$: YNO lower than SIDC), and the error metrics SEE, RMSE, and MAE measure prediction accuracy (smaller is better).

For the daily comparison ($n = 10,502$ matched days), the regression yields a slope of $0.9719 \pm 0.0032$ and an intercept of $0.267 \pm 0.026$, with $r = 0.9463$ ($R^2 = 0.8955$). The error metrics are SEE = 1.49, RMSE = 1.49, and MAE = 1.05. The slope of 0.972 indicates that YNO daily GN values are slightly higher than SIDC values. The 95\% confidence interval $[0.9656, 0.9783]$ does not include unity, confirming that this offset is statistically significant.

For the annual comparison ($n = 58$ years), the regression slope is $0.8997 \pm 0.0177$ with an intercept of $-0.056 \pm 0.120$, and $r = 0.9893$ ($R^2 = 0.9788$), with SEE = 0.50, RMSE = 0.49, and MAE = 0.34. Compared to the daily comparison, the annual analysis shows substantially higher correlation ($0.9893$ vs. $0.9463$) and much smaller error metrics (SEE: 0.50 vs. 1.49, MAE: 0.34 vs. 1.05), indicating that annual averaging reduces random scatter and provides better agreement with SIDC reference data. The slope of 0.900 indicates that annual mean YNO GN values are higher than SIDC annual values. The 95\% confidence interval $[0.8642, 0.9351]$ does not include unity, confirming the statistical significance of this offset.

\begin{table}[htbp]
\centering
\caption{Statistical summary YNO GN vs. SIDC comparison}
\label{tab:yno_comparison_stats_gn}
\begin{tabular}{lcccccccc}
\hline
Dataset & $n$ & $r$ & $R^2$ & Slope & Intercept & SEE & RMSE & MAE \\
\hline
YNO GN (daily) & 10,502 & 0.9463 & 0.8955 & $0.9719\pm0.0032$ & $0.267\pm0.026$ & 1.49 & 1.49 & 1.05 \\
YNO GN (annual) & 58 & 0.9893 & 0.9788 & $0.8997\pm0.0177$ & $-0.056\pm0.120$ & 0.50 & 0.49 & 0.34 \\
\hline
\end{tabular}
\end{table}
\subsubsection{Residual Analysis of GN for YNO}

\begin{figure}[htbp]
    \centering
    \includegraphics[width=\linewidth]{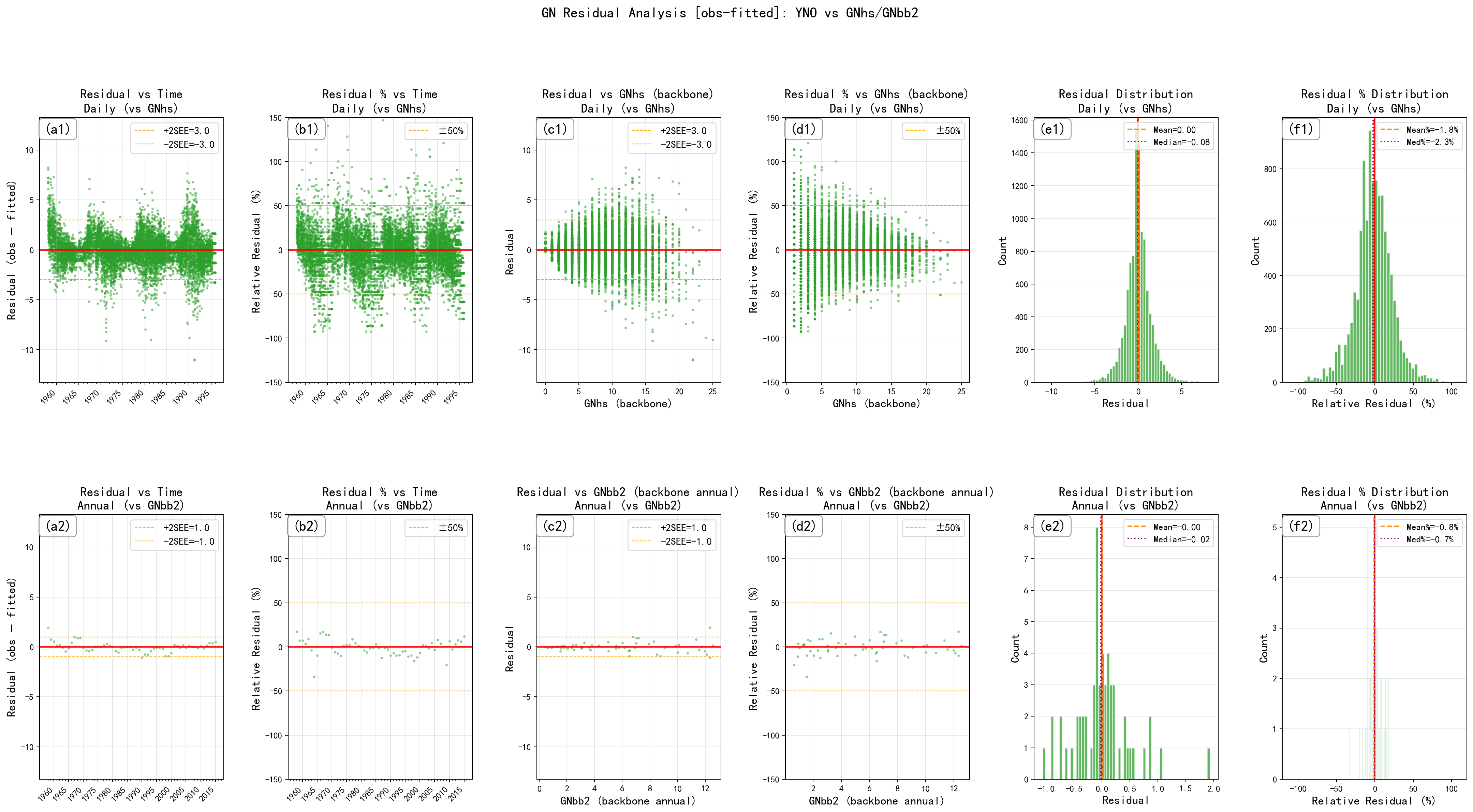}
    \caption{Residual analysis for YNO GN (obs $-$ fitted). Top row: daily comparison (vs GNhs, 1958--1995); bottom row: annual comparison (vs GNbb2, 1958--2015). Columns show (1) residuals vs time, (2) percentage residuals vs time, (3) residuals vs reference, (4) percentage residuals vs reference, (5) residual histogram, (6) percentage residual histogram.}
    \label{fig:gn_yno_residual_fitted}
\end{figure}

Figure~\ref{fig:gn_yno_residual_fitted} presents the residual analysis for YNO GN based on the regression model (residual = observed $-$ fitted). The figure consists of two rows (daily and annual comparisons) and six columns, following the same layout as Figure~\ref{fig:rsn_yno_residual_fitted}.

The first two columns (Col 1-2) show residuals and percentage residuals versus time. For daily YNO GN, residuals are scattered around zero with a standard error of the estimate (SEE) of 1.49. The $\pm2$SEE boundaries ($\pm2.99$) capture most data points. For annual data, the SEE reduces dramatically to 0.50, indicating that annual averaging significantly reduces scatter.

The absolute residuals (Col 1) for daily data show larger dispersion, while the annual residuals are much more tightly clustered. The percentage residuals (Col 2) exhibit larger relative errors at low activity levels, consistent with the behavior observed in the RSN analysis. The scatter plots (Col 3-4) show patterns consistent with the RSN analysis: absolute residuals increase with reference GN values (higher activity level leads to larger absolute deviations), while percentage residuals show larger scatter at low activity levels (relative errors are larger when activity is low).

The histograms (Col 5-6) show the distribution of residuals and percentage residuals. For daily YNO GN (Col 5), the residual distribution is centered near zero with a median of $-0.08$. The percentage residual histogram (Col 6) shows a mean of $-1.8\%$ and a median of $-2.3\%$. For the annual comparison (bottom row), the residual distribution shows even better concentration, with a median of $-0.02$, while the percentage residuals show a mean of $-0.8\%$ and a median of $-0.7\%$. The annual residuals are substantially more concentrated around zero compared to the daily residuals, with much smaller spread, confirming that annual averaging provides superior agreement with SIDC reference data for long-term trend analysis.

\subsection{PMO Sunspot Observation Comparison}

\subsubsection{Relative Sunspot Number (RSN) Analysis for PMO}
\begin{figure}[htbp]
    \centering
    \includegraphics[width=\linewidth]{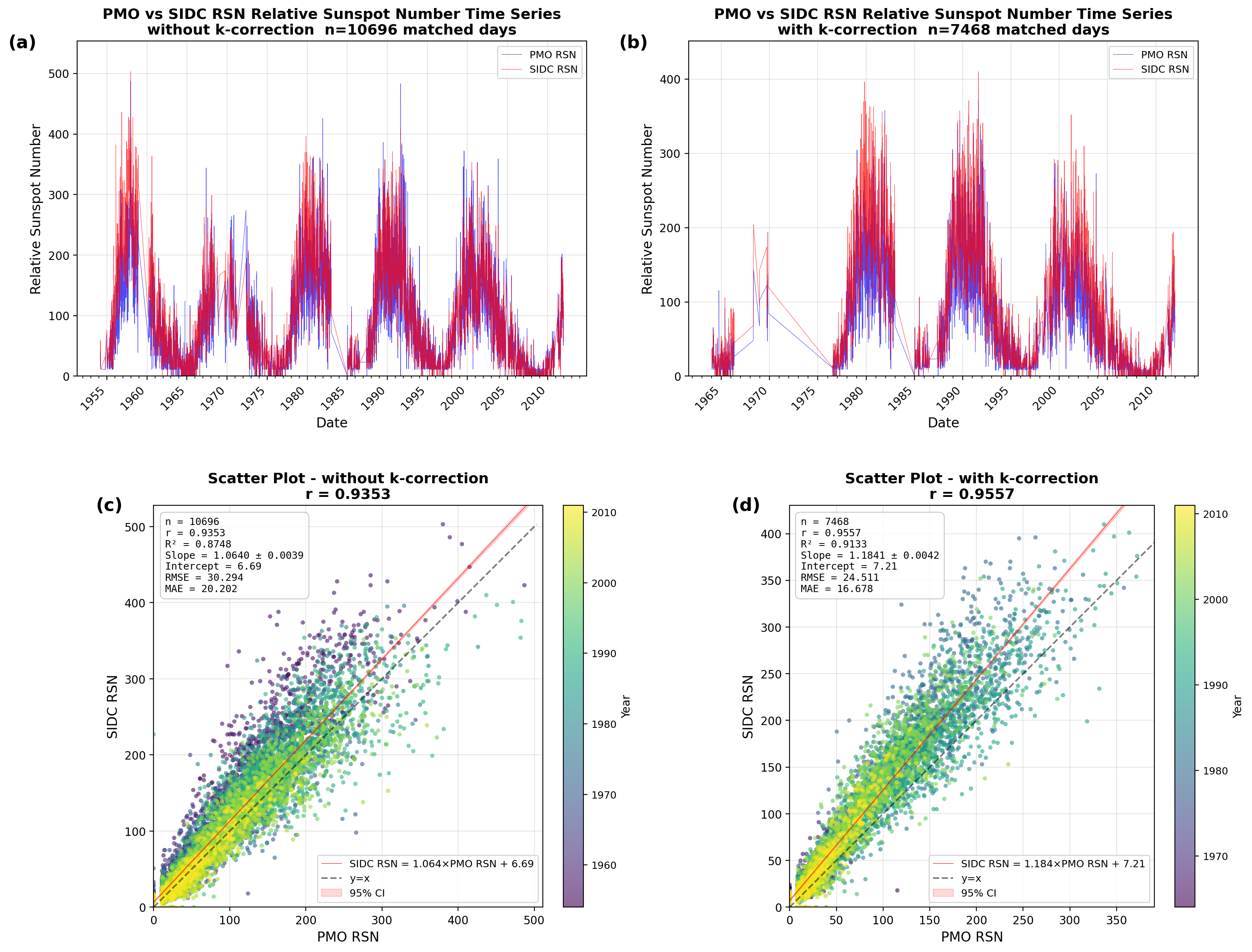}
    \caption{Comparison between PMO relative sunspot numbers (RSN) and SIDC reference data. The top row shows time series comparison; the bottom row shows scatter plots with regression analysis. Raw (left column) and k-corrected (right column) comparisons are presented. Legend boxes display statistical parameters including sample size $n$, correlation coefficient $r$, coefficient of determination $R^2$, regression slope and intercept, RMSE, and MAE.}
    \label{fig:PMO_SIDC_RSN}
\end{figure}

\begin{table}[htbp]
\centering
\caption{Statistical summary of PMO RSN vs. SIDC comparison}
\label{tab:pmo_comparison_stats_rsn}
\begin{tabular}{lcccccccc}
\hline
Dataset & $n$ & $r$ & $R^2$ & Slope & Intercept & SEE & RMSE & MAE \\
\hline
PMO RSN (raw) & 10,696 & 0.9353 & 0.8748 & $1.0640\pm0.0039$ & $6.694\pm0.451$ & 30.30 & 30.29 & 20.20 \\
PMO RSN (k) & 7,468 & 0.9557 & 0.9133 & $1.1841\pm0.0042$ & $7.211\pm0.422$ & 24.51 & 24.51 & 16.68 \\
\hline
\end{tabular}
\end{table}

Figure~\ref{fig:PMO_SIDC_RSN} presents the comparison between PMO relative sunspot numbers (RSN) and SIDC reference data, with the same layout as Figure~\ref{fig:YNO_SIDC_RSN}. The regression statistics for PMO RSN are compiled in Table~\ref{tab:pmo_comparison_stats_rsn}, following the same format as Table~\ref{tab:yno_comparison_stats_rsn}.

For the raw PMO RSN (without k-correction), regression against SIDC reference data yields a slope of $1.0640 \pm 0.0039$ and an intercept of $6.694 \pm 0.451$. The correlation reaches $r = 0.9353$ ($R^2 = 0.8748$), based on $n = 10,696$ daily matched pairs. The error statistics are SEE = 30.30, RMSE = 30.29, and MAE = 20.20. A slope of 1.064 implies that raw PMO RSN values tend to be lower than SIDC values. The 95\% confidence interval $[1.0564, 1.0717]$ excludes unity, confirming a statistically meaningful offset.

For the k-corrected PMO RSN ($n = 7,468$ matched days), the k-correction factors vary between 0.710 and 1.680, averaging 0.829. After correction, the regression slope becomes $1.1841 \pm 0.0042$ with an intercept of $7.211 \pm 0.422$. The correlation improves to $r = 0.9557$ ($R^2 = 0.9133$). All three error metrics decline: SEE falls from 30.30 to 24.51, RMSE from 30.29 to 24.51, and MAE from 20.20 to 16.68, demonstrating that k-correction enhances predictive performance. The post-correction slope of 1.184 shows that PMO RSN values remain systematically lower than SIDC. The 95\% confidence interval $[1.1758, 1.1923]$ again excludes unity, underscoring the statistical significance of this offset.

\subsubsection{Residual Analysis of RSN for PMO}

\begin{figure}[htbp]
    \centering
    \includegraphics[width=\linewidth]{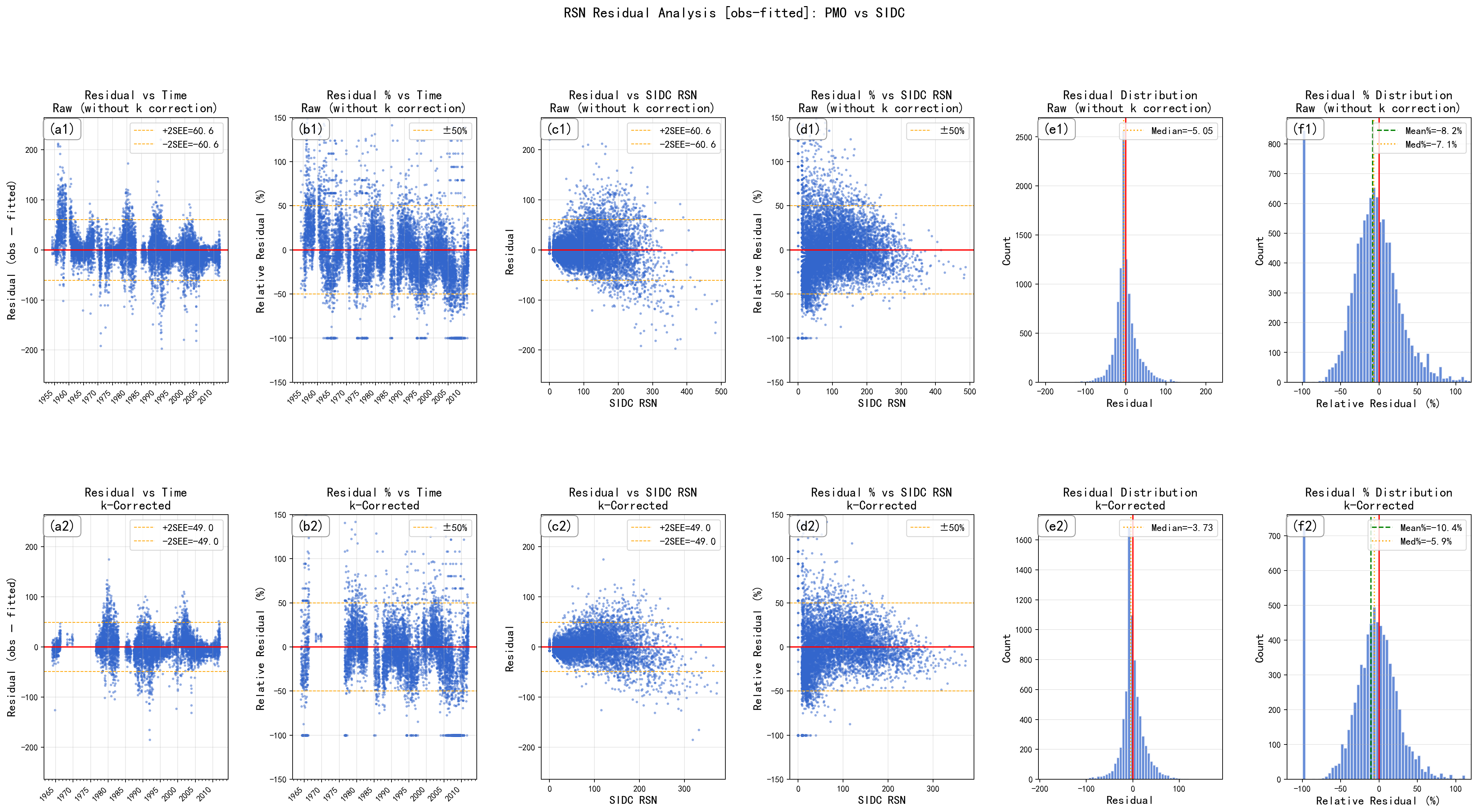}
    \caption{Residual analysis for PMO RSN (obs $-$ fitted). Top row: raw data; bottom row: k-corrected data. Columns show (1) residuals vs time, (2) percentage residuals vs time, (3) residuals vs reference SIDC RSN, (4) percentage residuals vs reference, (5) residual histogram, (6) percentage residual histogram. Dashed lines indicate $\pm2$SEE (orange) and $\pm50\%$ (orange).}
    \label{fig:rsn_pmo_residual_fitted}
\end{figure}

Figure~\ref{fig:rsn_pmo_residual_fitted} presents the residual analysis for PMO RSN based on the regression model (residual = observed $-$ fitted), following the same layout as Figure~\ref{fig:rsn_yno_residual_fitted}.

The first two columns (Col 1-2) show residuals and percentage residuals versus time. For raw PMO RSN, residuals are scattered around zero with a standard error of the estimate (SEE) of 30.30. The $\pm2$SEE boundaries ($\pm60.59$) capture most data points. For k-corrected data, the SEE reduces to 24.51, indicating improved precision after correction, with residuals more tightly clustered around zero.

The absolute residuals (Col 1) also exhibit a clear solar cycle modulation like those of Figure~\ref{fig:rsn_yno_residual_fitted}. The scatter plots (Col 3-4) show patterns consistent with the YNO analysis: absolute residuals increase with reference RSN values, while percentage residuals show larger scatter at low activity levels. After k-correction, both absolute and percentage residuals show reduced overall scatter, consistent with the improved SEE and RMSE values in Table~\ref{tab:pmo_comparison_stats_rsn}.

The histograms (Col 5-6) show the distribution of residuals and percentage residuals. For raw PMO RSN (Col 5), the residual distribution is centered near zero with a median of $-5.05$. For k-corrected data, the median improves to $-3.73$, consistent with the reduced RMSE (from 30.29 to 24.51). The percentage residual histograms (Col 6) show that for raw data, the mean percentage residual is $-8.2\%$ with a median of $-7.1\%$. After k-correction, the mean percentage residual becomes $-10.4\%$ with a median of $-5.9\%$, while the distribution shows a modest improvement in concentration around zero.
\subsubsection{Group Number (GN) Analysis for PMO}

\begin{figure}[htbp]
    \centering
    \includegraphics[width=\linewidth]{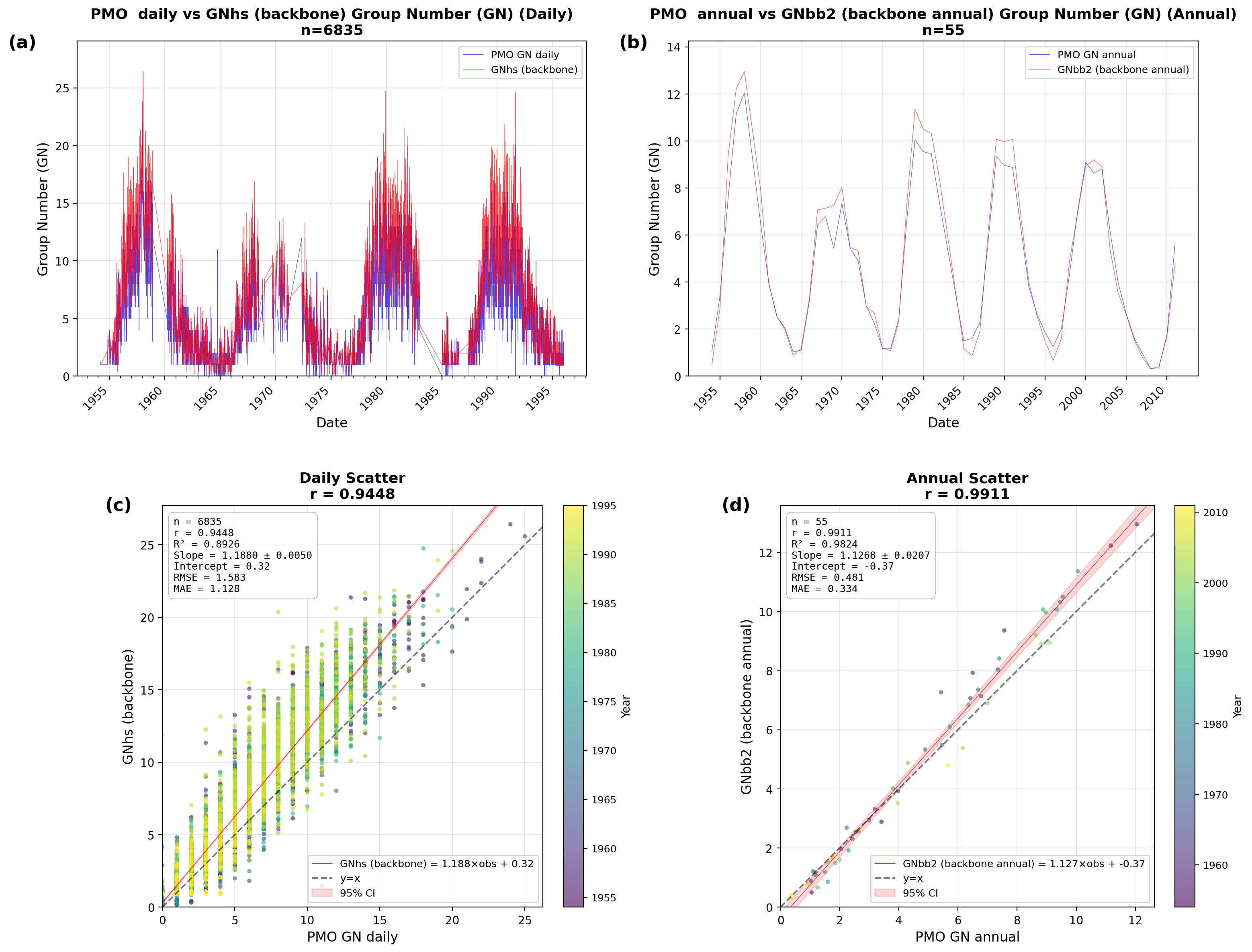}
    \caption{Comparison between PMO group numbers (GN) and SIDC reconstructed GN series (GNhs daily and GNbb2 annual). The top row shows time series comparisons; the bottom row shows scatter plots with regression analysis. Daily comparison (left column) uses GNhs\_d.txt (1958–1995); annual comparison (right column) uses GNbb2\_y.txt (1958–2015). Legend boxes display the same statistical parameters as in Figure~\ref{fig:PMO_SIDC_RSN}.}
    \label{fig:PMO_SIDC_GN}
\end{figure}

Figure~\ref{fig:PMO_SIDC_GN} illustrates the comparison for group numbers (GN), treating daily and annual analyses separately. The daily comparison (1958–1995) employs the SILSO reconstructed daily series (GNhs\_d.txt), while the annual comparison (1958–2015) uses the annual backbone series (GNbb2\_y.txt). Annual GN values for PMO are computed as yearly averages of daily observations.

The statistical outcomes for PMO GN are collected in Table~\ref{tab:pmo_comparison_stats_gn}. As in the RSN analysis, $r$ measures linear agreement, the slope indicates systematic offset (slope $<1$: PMO exceeds SIDC; slope $>1$: PMO falls below SIDC), and SEE, RMSE, and MAE quantify prediction accuracy (lower values signify better agreement).

For the daily comparison ($n = 6,835$ matched days), regression produces a slope of $1.1880 \pm 0.0050$ and an intercept of $0.316 \pm 0.034$, with $r = 0.9448$ ($R^2 = 0.8926$). The error metrics are SEE = 1.58, RMSE = 1.58, and MAE = 1.13. A slope of 1.188 reveals that PMO daily GN values are systematically lower than SIDC daily values. The 95\% confidence interval $[1.1782, 1.1978]$ excludes unity, confirming statistical significance.

For the annual comparison ($n = 55$ years), the regression yields a slope of $1.1268 \pm 0.0207$ and an intercept of $-0.373 \pm 0.119$, with $r = 0.9911$ ($R^2 = 0.9824$), SEE = 0.49, RMSE = 0.48, and MAE = 0.33. Relative to the daily analysis, the annual version exhibits markedly higher correlation ($0.9911$ vs. $0.9448$) and substantially reduced error metrics (SEE: 0.49 vs. 1.58, MAE: 0.33 vs. 1.13), demonstrating that annual averaging effectively suppresses random scatter and yields closer alignment with SIDC. The slope of 1.127 indicates that annual mean PMO GN values lie below SIDC annual values. The 95\% confidence interval $[1.0852, 1.1685]$ again excludes unity, solidifying the statistical significance of this offset.

\begin{table}[htbp]
\centering
\caption{Statistical summary PMO GN vs. SIDC comparison}
\label{tab:pmo_comparison_stats_gn}
\begin{tabular}{lcccccccc}
\hline
Dataset & $n$ & $r$ & $R^2$ & Slope & Intercept & SEE & RMSE & MAE \\
\hline
PMO GN (daily) & 6,835 & 0.9448 & 0.8926 & $1.1880\pm0.0050$ & $0.316\pm0.034$ & 1.58 & 1.58 & 1.13 \\
PMO GN (annual) & 55 & 0.9911 & 0.9824 & $1.1268\pm0.0207$ & $-0.373\pm0.119$ & 0.49 & 0.48 & 0.33 \\
\hline
\end{tabular}
\end{table}

\subsubsection{Residual Analysis of GN for PMO}

\begin{figure}[htbp]
    \centering
    \includegraphics[width=\linewidth]{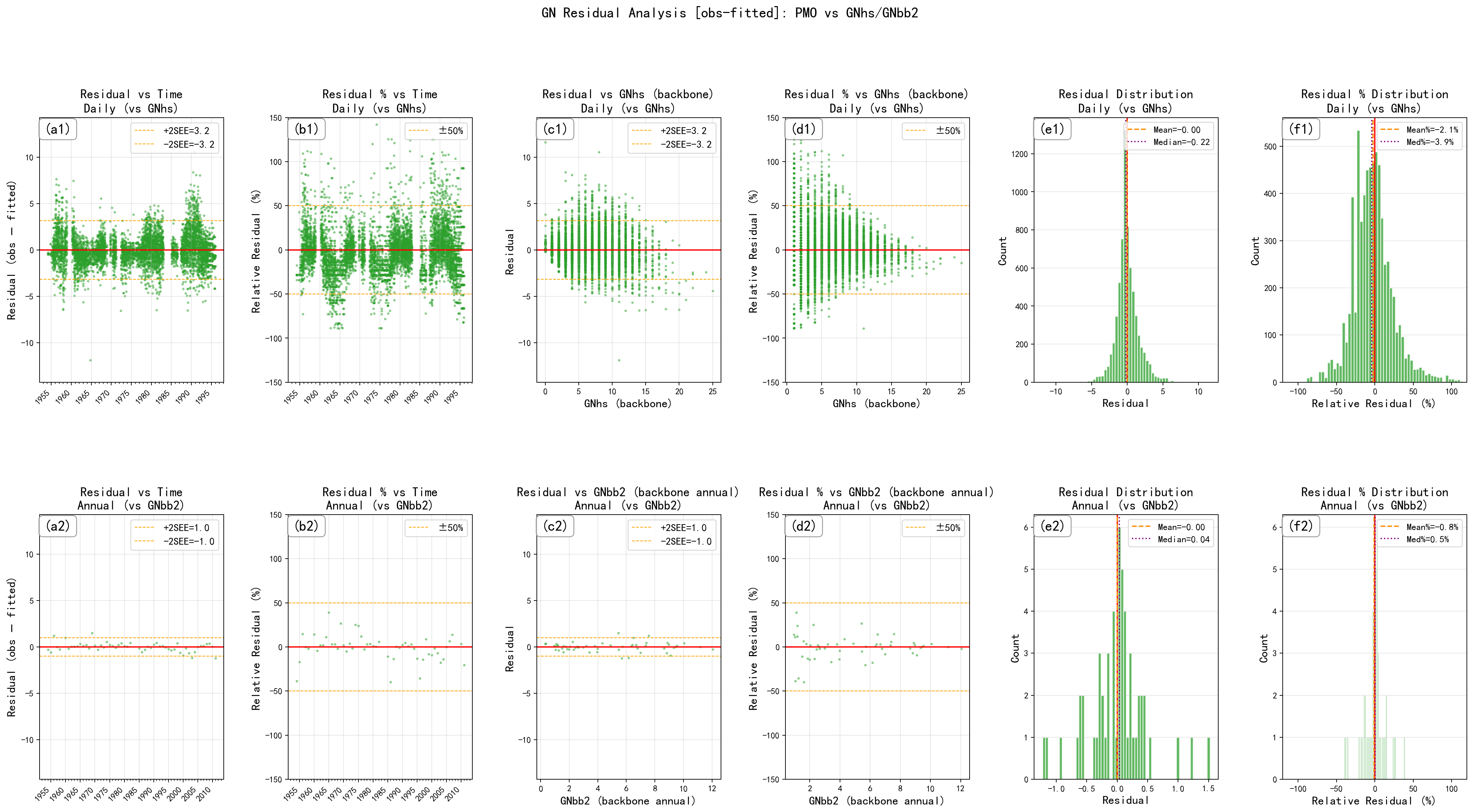}
    \caption{Residual analysis for PMO GN (obs $-$ fitted). Top row: daily comparison (vs GNhs, 1958–1995); bottom row: annual comparison (vs GNbb2, 1958–2015). Columns show (1) residuals vs time, (2) percentage residuals vs time, (3) residuals vs reference, (4) percentage residuals vs reference, (5) residual histogram, (6) percentage residual histogram.}
    \label{fig:gn_pmo_residual_fitted}
\end{figure}

Figure~\ref{fig:gn_pmo_residual_fitted} presents the residual analysis for PMO GN based on the regression model (residual = observed $-$ fitted). The figure consists of two rows (daily and annual comparisons) and six columns, following the same layout as Figure~\ref{fig:gn_yno_residual_fitted}.

The first two columns (Col 1-2) show residuals and percentage residuals versus time. For daily PMO GN, residuals are scattered around zero with a standard error of the estimate (SEE) of 1.58. The $\pm2$SEE boundaries ($\pm3.17$) capture most data points. For annual data, the SEE reduces dramatically to 0.49, indicating that annual averaging significantly reduces scatter.

The absolute residuals (Col 1) for daily data show larger dispersion, while the annual residuals are much more tightly clustered. The percentage residuals (Col 2) exhibit larger relative errors at low activity levels, consistent with the behavior observed in the YNO GN analysis. The scatter plots (Col 3-4) show patterns consistent with the YNO GN analysis: absolute residuals increase with reference GN values, while percentage residuals show larger scatter at low activity levels.

The histograms (Col 5-6) show the distribution of residuals and percentage residuals. For daily PMO GN (Col 5), the residual distribution is centered near zero with a median of $-0.22$. The percentage residual histogram (Col 6) shows a mean of $-2.1\%$ and a median of $-3.9\%$. For the annual comparison (bottom row), the residual distribution shows even better concentration, with a median of $0.04$, while the percentage residuals show a mean of $-0.8\%$ and a median of $0.5\%$. The annual residuals are substantially more concentrated around zero compared to the daily residuals, with much smaller spread, confirming that annual averaging provides superior agreement with SIDC reference data for long-term trend analysis, as also observed for YNO.

\section{Conclusion and Discussion}
\label{sec:disc}

In this work, we have presented a systematically revised and manually verified digital catalog of historical sunspot drawing observations from Chinese observatories, encompassing sunspot number, sunspot group number, and sunspot area measurements. Through comprehensive comparisons with international reference data from the SIDC, we have evaluated the quality and consistency of the Chinese sunspot records, focusing on Yunnan Observatory (YNO) and Purple Mountain Observatory (PMO).

\subsection*{Summary of Main Results}

The comparative analyses reveal consistently high correlations between the Chinese observations and SIDC reference data, with correlation coefficients typically exceeding 0.94 for all comparisons and reaching as high as 0.99 for annual GN averages (Table~\ref{tab:correlation_summary}).

For YNO RSN, the raw data yield a slope of $0.8497 \pm 0.0022$ ($r = 0.9517$), indicating uncorrected YNO values are systematically higher than SIDC by about 17.6\%. After k-correction (k factors: 0.510-0.980, mean 0.656), the slope becomes $1.2641 \pm 0.0031$ ($r = 0.9598$), shifting to a 20.9\% deficit. Error metrics improve consistently: SEE decreases from 25.03 to 21.79, RMSE from 25.03 to 21.79, and MAE from 16.60 to 14.81.

For PMO RSN, raw data show a slope of $1.0640 \pm 0.0039$ ($r = 0.9353$), meaning uncorrected PMO values are 6.0\% lower than SIDC. After k-correction (k factors: 0.710-1.680, mean 0.829), the slope becomes $1.1841 \pm 0.0042$ ($r = 0.9557$), with PMO values remaining 15.5\% lower. Error metrics also improve: SEE from 30.30 to 24.51, RMSE from 30.29 to 24.51, and MAE from 20.20 to 16.68.

For GN comparisons, daily YNO GN yields a slope of $0.9719 \pm 0.0032$ ($r = 0.9463$), while annual YNO GN shows $0.8997 \pm 0.0177$ ($r = 0.9893$). For PMO GN, daily data give $1.1880 \pm 0.0050$ ($r = 0.9448$), and annual data give $1.1268 \pm 0.0207$ ($r = 0.9911$). Annual averaging dramatically reduces scatter, with SEE dropping from 1.49 to 0.50 for YNO and from 1.58 to 0.49 for PMO.

The residual analyses confirm these findings. Absolute residuals exhibit clear solar cycle modulation — larger during maxima and smaller during minima — as expected. Percentage residuals show the opposite behavior, with larger relative errors during minima when fitted values are small. After k-correction, both absolute and percentage residuals show reduced scatter, with SEE improving from 25.03 to 21.79 for YNO RSN and from 30.30 to 24.51 for PMO RSN.

\subsection*{Key Observations}

A notable difference emerges between the two observatories. YNO raw RSN values are higher than SIDC (slope 0.850), while PMO raw RSN values are lower than SIDC (slope 1.064). After k-correction, YNO shifts from higher to lower (slope 1.264), suggesting over-correction, while PMO remains lower with a larger offset (slope 1.184), suggesting under-correction. These differing behaviors likely reflect distinct observational practices and calibration procedures at the two stations.

Consistent with the principles of the Group Sunspot Number \citep{1998SoPh..179..189H, 1998SoPh..181..491H}, our analysis confirms that the group number is a more stable parameter than the relative sunspot number. The systematic offsets for GN are smaller than those for RSN: for YNO, the RSN deviation is -20.9\% (after k-correction) while the GN deviation is -16.5\% (raw); for PMO, the RSN deviation is -15.5\% (after k-correction) while the GN deviation is -8.5\% (raw). This makes the group number a particularly robust index for long-term solar cycle studies, as it requires no additional correction factors and exhibits smaller systematic biases.

\subsection*{Limitations and Future Work}

Several limitations should be acknowledged. First, k-correction factors for PMO are only partially available, particularly in the early years. Second, the daily GN comparison is limited to 1995 due to the end of the SILSO reconstructed daily series. Third, the subjective nature of manual sunspot drawing inevitably introduces observer-dependent variability.

Future work should focus on developing advanced image processing techniques to extract high-precision area information from scanned drawings, and on establishing cross-calibration methods for area measurements across different stations. The potential recovery of early Qingdao Observatory records (1925–1947) would extend temporal coverage to a full century, providing an unprecedented long-term record from China.

\subsection*{Concluding Remarks}

In conclusion, the Chinese historical sunspot drawing archive, through systematic digitization, rigorous quality control, and careful calibration, provides a valuable and reliable resource for solar physics research. The strong correlations with SIDC data confirm that these visual observations faithfully capture solar activity cycles over multiple decades. This dataset serves as a unique eastern hemisphere complement to the predominantly western-dominated international sunspot records.

\begin{table}[htbp]
    \centering
    \caption{Summary of correlation statistics between Chinese observatories and SIDC reference data. For GN, daily comparison uses GNhs\_d.txt (1958--1995), and annual comparison uses GNbb2\_y.txt (1958--2015).}
    \label{tab:correlation_summary}
    \begin{tabular}{lcccccccc}
        \hline
        \textbf{Station} & \textbf{Parameter} & \textbf{n (raw)} & \textbf{n (k-corr)} & \textbf{r (raw)} & \textbf{r (k-corr)} & \textbf{Slope (raw)} & \textbf{Slope (k-corr)} & \textbf{k range (mean)} \\
        \hline
        YNO & RSN & 15,294 & 14,541 & 0.9517 & 0.9598 & 0.8497 & 1.2641 & 0.510-0.980 (0.656) \\
        YNO & GN (daily) & 10,502 & -- & 0.9463 & -- & 0.9719 & -- & -- \\
        YNO & GN (annual) & 58 & -- & 0.9893 & -- & 0.8997 & -- & -- \\
        PMO & RSN & 10,696 & 7,468 & 0.9353 & 0.9557 & 1.0640 & 1.1841 & 0.710-1.680 (0.829) \\
        PMO & GN (daily) & 6,835 & -- & 0.9448 & -- & 1.1880 & -- & -- \\
        PMO & GN (annual) & 55 & -- & 0.9911 & -- & 1.1268 & -- & -- \\
        \hline
    \end{tabular}
\end{table}

\begin{acknowledgments}
We sincerely thank the anonymous referee for the valuable suggestions and constructive comments, which have significantly improved the overall quality of this paper.
We are deeply grateful to Professor Deng Linhua for his dedicated professional guidance and valuable suggestions. We thank Professor Ji Haisheng from Purple Mountain Observatory for his work in compiling and digitizing the early multi-station data from PMO, and Professor Lin Jun from Yunnan Observatory for his work in compiling and digitizing the YNO data. We also extend our appreciation to all colleagues who participated in the digitization of Chinese hand-drawn sunspot drawings, quality verification, and further discussions on solar cycle studies using the digitized data. Special thanks go to the predecessors who devoted themselves to sunspot observation records over the past century, and we express our sincere respect for their contributions. 
This work was supported by the National Key R\&D Program of China (2021YFA1600500 and 2022YFF0503001), China's Space Origins Exploration Program (GJ11020400) and the Basic Work Special Project of the Ministry of Science and Technology of China (Grant No. 2014FY120300).
\end{acknowledgments}

\subsection*{Data Availability}
The digitized sunspot drawing parameters from all six Chinese observatories are publicly available at \url{https://sun10.bao.ac.cn/hsos_data/SHDA/sd_info/}. This repository contains unified machine-readable catalogs, accompanied by a readMe file documenting and revision history. The dataset is also archived at Zenodo with the DOI \dataset{10.5281/zenodo.20455752}.

\begin{contribution}
All authors contributed equally to the paper.
\end{contribution}

\bibliography{lius}{}
\bibliographystyle{aasjournalv7}

\end{document}